\documentclass[a4paper]{jpconf}

\usepackage{citesort,iopams}
\usepackage{graphicx}

\newcommand{\bd}{\begin{displaymath}}
\newcommand{\ed}{\end{displaymath}}
\newcommand{\be}{\begin{equation}}
\newcommand{\ee}{\end{equation}}
\newcommand{\ba}{\begin{eqnarray}}
\newcommand{\ea}{\end{eqnarray}}

\begin{document}

\title{Quantumness beyond quantum mechanics}

\author{\'Angel S. Sanz}

\address{Instituto de F{\'\i}sica Fundamental (IFF--CSIC),
Serrano 123, 28006 - Madrid, Spain}

\ead{asanz@iff.csic.es}

\begin{abstract}
Bohmian mechanics allows us to understand quantum systems in the
light of other quantum traits than the well-known ones (coherence,
diffraction, interference, tunneling, discreteness, entanglement,
etc.).
Here the discussion focusses precisely on two of these interesting
aspects, which arise when quantum mechanics is though within this
theoretical framework: the {\it non-crossing property}, which allows
for distinguishability without erasing interference patterns, and the
possibility to define {\it quantum probability tubes}, along which the
probability remains constant all the way.
Furthermore, taking into account this hydrodynamic-like description as
a link, it is also shown how this knowledge (concepts and ideas) can be
straightforwardly transferred to other fields of physics (for example,
the transmission of light along waveguides).
\end{abstract}

%%%%%%%%%%%%%%%%%%%%%%%%%%%%%%%%%%%%%%%%%%%%%%%%%%%%%%%%%%%%%%%%%%%%%%%
%%%%%%%%%%%%%%%%%%%%%%%%%%%%%%%%%%%%%%%%%%%%%%%%%%%%%%%%%%%%%%%%%%%%%%%

\section{Introduction}
\label{sec1}

At present, there is no doubt that quantum mechanics can be considered
the most successful theory ever devised to explain the physical world.
Its applicability ranges from very fundamental physical problems to the
high-technology applications that are nowadays an important part of our
daily life.
This theory, though, still constitutes a veiled mystery at a deeper
level of understanding, for there is a lack of a clear interpretation
of the physics underlying quantum systems.
This is somehow connected to its widely accepted interpretation, namely
the Copenhagen interpretation \cite{landsman:CQP:2009}, which not only
does not allow us to think of quantum systems as we do of classical
ones, but it just forbids such a thing.

A feasible way to surmount this drawback (although surely not the final
one) comes through Bohmian mechanics \cite{bohm:PR:1952-1,bohm:PR:1952-2,%
holland-bk,bohm-hiley-bk}.
According to this approach to the quantum world, systems are described
in terms of trajectories evolving in configuration space.
Rather than constituting a step backwards towards the classical
paradigm, this quantum-mechanical formulation puts at the same
descriptive level both classical and quantum systems, though the
latter's behavior differs from the former's one through the dynamical
particularities induced by the quantum laws of motion.
For instance, these laws give rise to motions constrained to the
configuration space, uncertainty principles, interference phenomena,
discreteness, and other features which are absent in classical systems.
Of course, in the statistical limit, i.e., when an appropriate sampling
over many of such Bohmian trajectories is carried out, the results of
the standard quantum mechanics are recovered.

On the other hand, we find that in principle there are no restrictions
to export the core idea in Bohmian mechanics, namely the possibility to
reformulate quantum mechanics as a hydrodynamic-like theory, as well as
the ideas emerging from it to other physical contexts where waves or
distributions are the primary descriptor.
This is somehow the other way around that Madelung
\cite{madelung:ZPhys:1926} considered in 1926, when he tried to provide
a hydrodynamical interpretation of quantum processes based on the
analogy between the latter and classical hydrodynamical flows.
Within this approach, Bohmian trajectories were just considered to be
the ``tracks'' displayed by some hypothetical {\it tracer particles}
left on top of the quantum fluid \cite{sanz:AJP:2012}, something very
similar to what can be found in modern studies of flows, where the
trajectories displayed by a number of tracer particles is recorded and
subsequently analyzed in order to extract information about the flow.
By means of this method, for example, studies revealed anomalous
diffusion and L\'evy flights in two-dimensional rotating flows
\cite{swinney:PRL:1993,klafter:PhysToday:1996}.
However, closer in spirit to the present discussion, we find the recent
observations by Couder and coworkers
\cite{couder:Nature:2005,couder:JFluidMech:2006,couder:PRL:2006,couder:PRL:2009,couder:PNAS:2010}
as well as by Bush \cite{bush:PNAS:2010}, where a seemingly classical
analog of Bohmian motion is found for droplets bouncing on a vertically
vibrating bath of the same fluid.

The purpose of the present work is to pose an analysis and discussion
of some interesting properties of quantum systems and their
consequences, which arise when they are investigated from a Bohmian
perspective.
Furthermore, the link that can be established between the Bohmian
viewpoint and its application or extension to other non-quantum wave
theories will also be discussed.
In order to make clear the arguments exposed, the work has been
organized as follows.
In Section~\ref{sec2}, the essential elements/concepts of Bohmian
mechanics are introduced as well as some important consequences
implied by this theory.
The transfer of knowledge from Bohmian mechanics to other classical
wave theories, such as electromagnetism, is discussed in Section
\ref{sec3} in the context of light waveguiding.
To conclude, a series of final remarks are highlighted in
Section~\ref{sec4}.

%%%%%%%%%%%%%%%%%%%%%%%%%%%%%%%%%%%%%%%%%%%%%%%%%%%%%%%%%%%%%%%%%%%%%%%
%%%%%%%%%%%%%%%%%%%%%%%%%%%%%%%%%%%%%%%%%%%%%%%%%%%%%%%%%%%%%%%%%%%%%%%

\section{What can be learnt from Bohmian mechanics?}
\label{sec2}

%%%%%%%%%%%%%%%%%%%%%%%%%%%%%%%%%%%%%%%%%%%%%%%%%%%%%%%%%%%%%%%%%%%%%%%

\subsection{Some elementary background}
\label{sec2-1}

In the standard Bohmian approach \cite{bohm:PR:1952-1}, one usually
starts considering the wave function in polar form,
\begin{equation}
 \Psi({\bf r},t) = R({\bf r},t) e^{iS({\bf r},t)/\hbar} ,
 \label{e2}
\end{equation}
which is substituted into the (single-particle) non-relativistic,
time-dependent Schr\"odinger equation,
\begin{equation}
 i\hbar\ \frac{\partial \Psi}{\partial t} =
  \left( - \frac{\hbar^2}{2m}\ \nabla^2 + V \right) \Psi ,
 \label{e1}
\end{equation}
to yield the system of real coupled equations
\setlength\arraycolsep{2pt}
\begin{eqnarray}
 \frac{\partial R^2}{\partial t} & + &
  \nabla \cdot \left( R^2\ \frac{\nabla S}{m} \right) = 0 ,
 \label{e3} \\
 \frac{\partial S}{\partial t} & + & \frac{(\nabla S)^2}{2m} +
  V - \frac{\hbar^2}{2m} \frac{\nabla^2 R}{R} = 0 .
 \label{e4}
\end{eqnarray}
In brief, the continuity (or conservation) equation (\ref{e3}) rules
the ensemble dynamics, i.e., the number of particles described by the
probability density has to remain constant; the quantum Hamilton-Jacobi
equation (\ref{e4}), on the other hand, describes the time-evolution of
the phase field, which governs the motion of quantum particles through
the guidance equation
\begin{equation}
 {\bf v}_B = \dot{\bf r} = \frac{\nabla S}{m} .
 \label{e10}
\end{equation}
In this way, the simple transformation described by (\ref{e2}) allows
us to go a step beyond standard quantum mechanics.
In the latter, $\Psi$ is all what is needed to describe and interpret
the experimental outcomes; in Bohmian mechanics, on the other hand,
the evolution of a {\it tracer particle} \cite{sanz:AJP:2012} according
to the information supplied by $\Psi$ (the guiding field) is also
considered to further understand the quantum system dynamics.

Now, the same description can be formally inferred by simply assuming
that the evolution of any quantum system can be identified with the
diffusion of a (quantum) fluid throughout configuration space, as
formerly proposed by Madelung \cite{madelung:ZPhys:1926}.
Therefore, as in the case of classical fluids, the elementary
quantities of this approach would be the (probability) density of
the quantum fluid and its associated probability current density,
${\bf J}$, both related through the conservation (or continuity)
equation
\begin{equation}
 \frac{\partial \rho}{\partial t} + \nabla \cdot {\bf J} = 0 ,
 \label{e9}
\end{equation}
where ${\bf J} = \rho {\bf v}_H$ and ${\bf v}_H$ is a local hydrodynamic
velocity field.
The probability density and the probability current density are related
to the wave function (\ref{e2}) by the transformation relations
\be
 \rho = R^2 , \qquad
 {\bf J} = R^2\ \frac{\nabla S}{m} ,
 \label{e8}
\ee
with the local velocity field being
\begin{equation}
 {\bf v}_H = \frac{{\bf J}}{\rho} = \frac{\nabla S}{m} ,
 \label{e10b}
\end{equation}
which is equivalent to the Bohmian one, ${\bf v}_B$, given by
(\ref{e10}) (hence, from now on, both velocity fields will be denoted
by ${\bf v}$).
The integration in time of (\ref{e10b}) generates a family of
streamlines or paths (for each wave function considered) along which
the quantum fluid propagates, just as in the case of a classical fluid.
As it can be inferred, the first equality goes beyond Bohmian mechanics
and allows to define streamlines in any system characterized by
a certain density and a vector that transports it through the
corresponding configuration space, regardless whether such a
density describes a quantum system or not.

%%%%%%%%%%%%%%%%%%%%%%%%%%%%%%%%%%%%%%%%%%%%%%%%%%%%%%%%%%%%%%%%%%%%%%%

\subsection{The non-crossing rule}
\label{sec2-2}

One of the most relevant properties revealed by Bohmian mechanics
is that quantum fluxes cannot cross in configuration space.
That is, at a certain time $t$, each point on configuration space has
a uniquely-defined value of the local velocity field, except for
regions where the wave function displays a node (i.e., $\Psi = 0$).
Therefore, two or more Bohmian trajectories cannot cross through such
a point at the same time $t$ or, equivalently, {\it quantum systems
cannot reach the same configuration at a given time following different
quantum streamlines}.
In terms of the standard quantum mechanics, this just means that paths
can be labeled, although they cannot be experimentally {\it observed}
because this implies changing the process and therefore generating
new, alternative paths.
Obviously, this changes dramatically our physical view of the
superposition principle, which states that any linear combination of
solutions of Schr\"odinger's equation is also a formally valid solution
to the same equation.
Accordingly, given a coherent superposition of two counter-propagating
wave packets, they can be evolved in time separately and then
recombined at any time to determine when and where an interference
pattern.
Formally, there is no problem in proceeding in this way.
However, from a (hydro) dynamical viewpoint the evolution of both wave
packets has to be accounted for together for the interference patter to
be observable (which is what actually happens in an experiment), this
implying that interpretations of physical phenomena cannot directly
rely upon the superposition principle.
Bohmian mechanics provides us an answer for this based precisely on
the non-crossing rule.

In order to make clear such a statement, consider the wave packets
above are represented by two free expanding Gaussian wave packets,
\begin{equation}
 \Psi(x,t) = \left( \frac{1}{2\pi\tilde{\sigma}_t^2} \right)^{1/4}
  e^{-(x - x_{\rm cl})^2/4\tilde{\sigma}_t\sigma_0
  + i p_0 (x - x_{\rm cl})/\hbar + iEt/\hbar} ,
 \label{e11}
\end{equation}
where the time-dependence of their spreading is given by
\begin{equation}
 \sigma_t = |\tilde{\sigma}_t| =
  \sigma_0 \sqrt{1 + \left( \frac{\hbar t}{2m\sigma_0^2} \right)^2} .
 \label{e12}
\end{equation}
The wave packet (\ref{e11}) propagates along the classical trajectory
$x_{\rm cl} = x_0 + v_0 t$ (with $v_0 = p_0/m$), while the associated
Bohmian trajectories are given by
\begin{equation}
 x(t) = x_{\rm cl} + \frac{\sigma_t}{\sigma_0}\
  \left[x(0) - x_0\right] ,
 \label{e13}
\end{equation}
with $x(0)$ denoting the initial position of the corresponding
Bohmian trajectory.
As can be noticed, there are two contributions to the evolution of the
tracer particles, one coming from the classical drift and another one
from a purely quantum origin \cite{sanz:cpl:2007,sanz:JPA:2008}.
The former makes the particle to evolve according to only the laws of
classical mechanics; the latter makes it to deviate in such a way that
its motion accommodates to the evolution of the quantum fluid.
The time-scale that separates both types of motion (to some extent)
is given in this particular case by the characteristic time
$\tau \equiv 2m\sigma_0^2/\hbar$ \cite{sanz:cpl:2007,sanz:JPA:2008}.
Depending on the relative value of $t$ when compared with $\tau$, three
regimes can then be found (in analogy with the regimes of optical
physics \cite{sanz:AJP:2012}):
\begin{itemize}
 \item {\it Ehrenfest or Huygens} ($\tau \equiv 2m\sigma_0^2/\hbar
  \ggg t$): Bohmian trajectories follow classical ones (in agreement
  with Ehrenfest's theorem).
 \item {\it Near-field or Fresnel} ($\tau \equiv 2m\sigma_0^2/\hbar
  > t$): Bohmian trajectories start to diverge from their classical
  counterparts.
 \item {\it Far-field or Franhofer} ($\tau \equiv 2m\sigma_0^2/\hbar
  \ll t$): Bohmian trajectories display a sort of rectilinear,
  uniform evolution due to the stationarity of the wave function
  far from any interaction potential.
\end{itemize}
Therefore, by inspecting the topology displayed by the quantum
trajectories, one readily notices that the quantum flow evolves from an
initially confined fluid to a linearly expanding one, undergoing at
times of the order of $\tau$ a sort of internal boosting which bursts
it open.
However, this does not provide any clue on the true dynamics.

\begin{figure}[t]
\centering
 \begin{minipage}{11pc}
  \includegraphics[width=10.25pc]{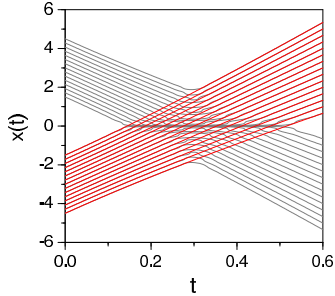}
 \end{minipage}\hspace{1.75pc}
 \begin{minipage}{21pc}
  \caption{\label{fig1}
   Bohmian trajectories associated with a coherent superposition of two
   Gaussian wave packets (gray lines) and with a single Gaussian wave
   packet (red lines).}
 \end{minipage}
\end{figure}

\begin{figure}[t]
 \centering
 \begin{minipage}{34pc}
  \includegraphics[width=10pc]{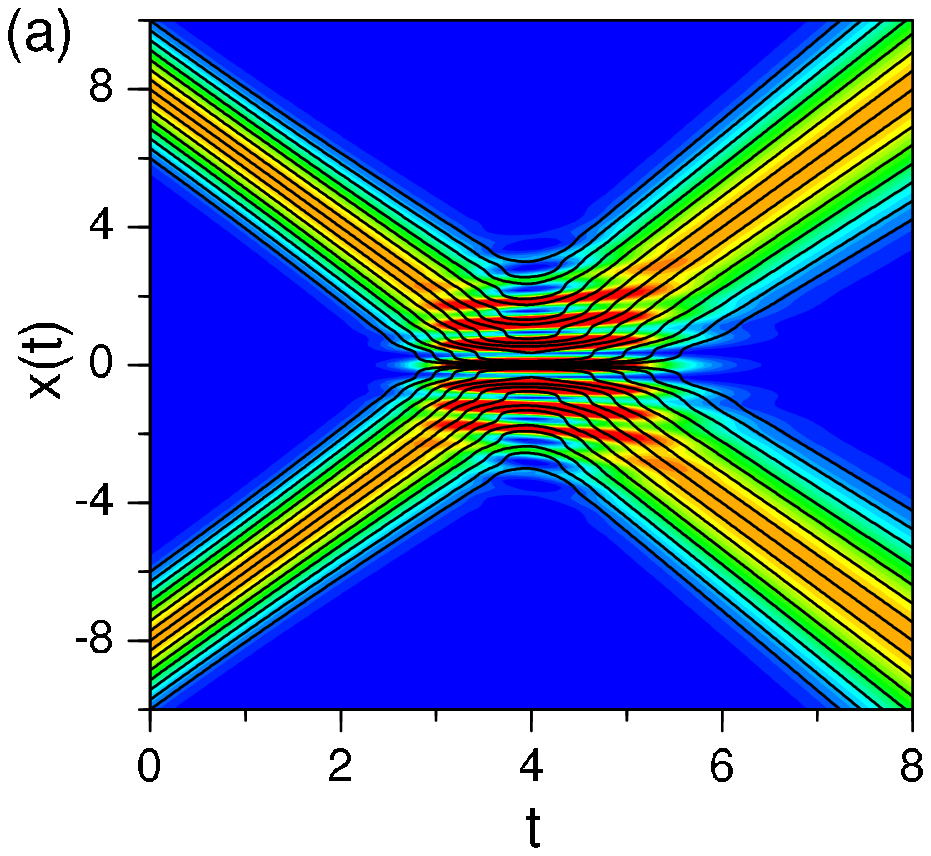}\hspace{1.5pc}
  \includegraphics[width=10pc]{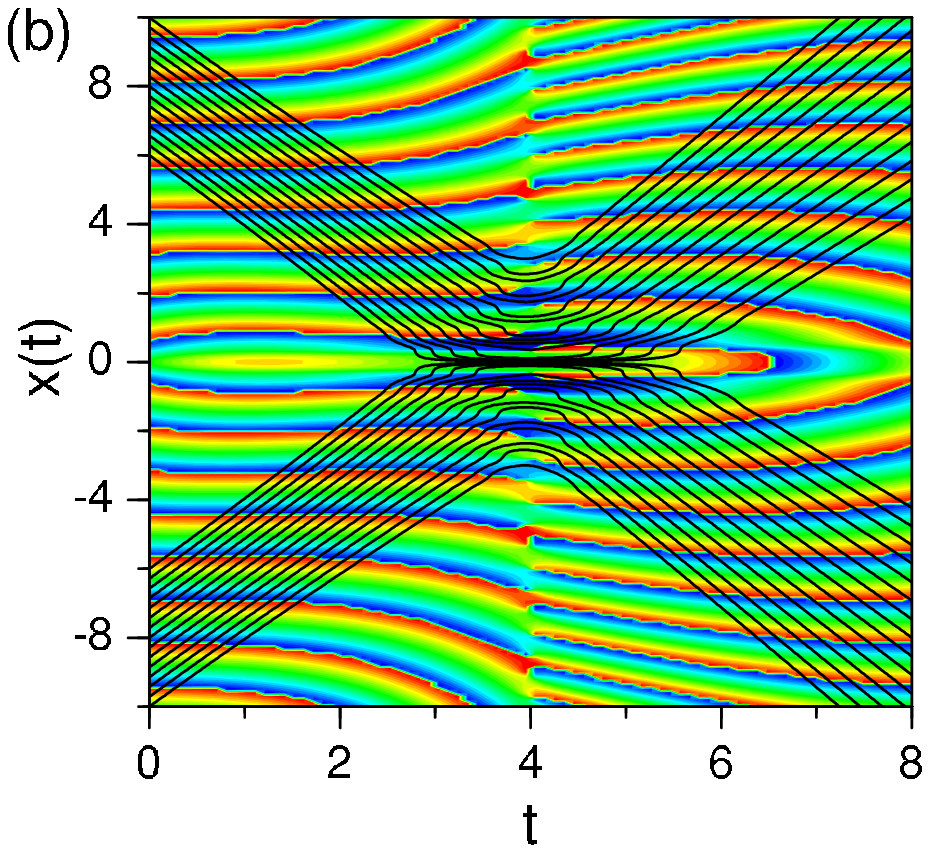}\hspace{1.5pc}
  \includegraphics[width=10pc]{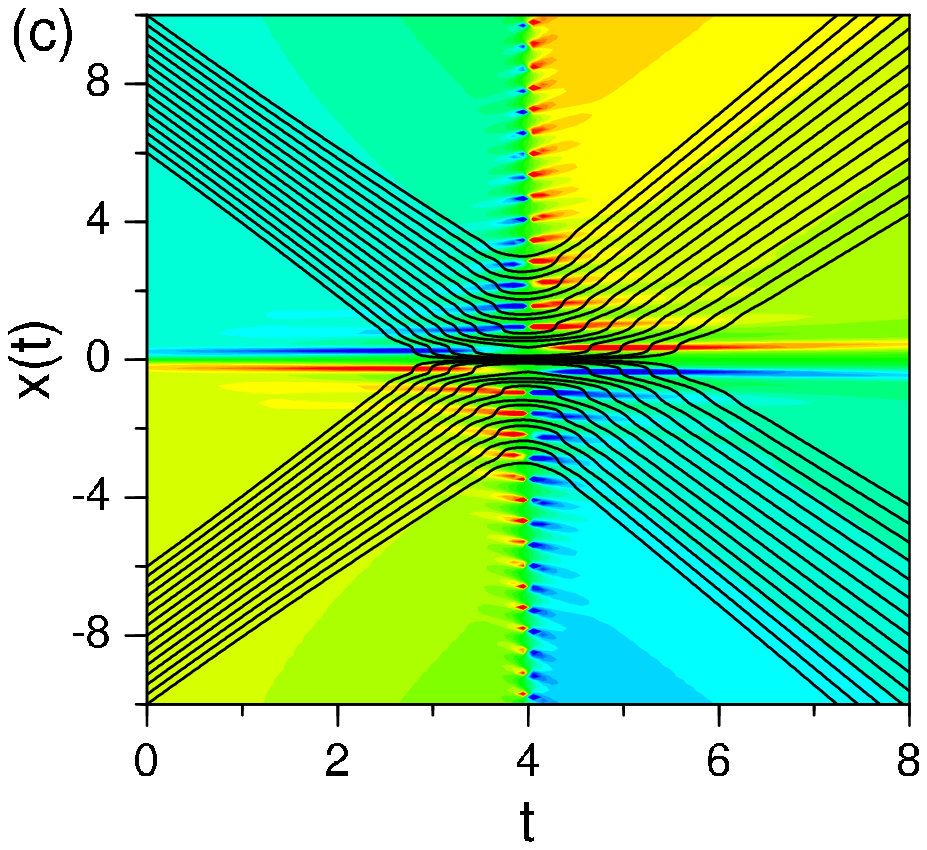}
 \end{minipage}
  \caption{\label{fig2}
   Contour-plots representing the evolution in time of the probability
   density (a), the phase field (b) and the local velocity field (c) of
   a initial coherent superposition of two Gaussian wave packets.
   The Bohmian trajectories (black lines) associated with this problem
   are also displayed superimposed to each graph.}
\end{figure}

In order to understand the importance of not disentangling the joint
evolution of both wave packets, as indicated by the superposition
principle, consider the full wave function,
\begin{equation}
 \Psi({\bf r},t) = \psi_1({\bf r},t) + \psi_2({\bf r},t) ,
 \label{e19}
\end{equation}
where the $\psi_i = \rho_i^{1/2} e^{iS_i/\hbar}$ are properly
normalized and have the shape indicated by (\ref{e11}), with
$x_{2,0} = - x_{1,0} = x_0 > 0$ and $v_{1,0} = - v_{2,0} = v_0 > 0$.
It is straightforward to notice that
\setlength\arraycolsep{2pt}
\begin{eqnarray}
 \rho & = & \rho_1 + \rho_2 + 2 \sqrt{\rho_1 \rho_2} \cos \varphi ,
 \label{e20} \\
 {\bf J} & = &
   \frac{1}{m} \left[ \rho_1 \nabla S_1 + \rho_2 \nabla S_2
    + \sqrt{\rho_1 \rho_2} \nabla (S_1 + S_2) \cos \varphi
    + \hbar \left( \rho_1^{1/2} \nabla \rho_2^{1/2}
      - \rho_2^{1/2} \nabla \rho_1^{1/2} \right) \sin \varphi \right] ,
 \label{e21}
\end{eqnarray}
with $\varphi = (S_2 - S_1)/\hbar$.
Substituting these expressions into (\ref{e10}) lead to the equation
of motion
\be
 \dot{\bf r} = \frac{1}{m}
  \frac{\rho_1 \nabla S_1 + \rho_2 \nabla S_2
  + \sqrt{\rho_1 \rho_2} \nabla (S_1 + S_2) \cos \varphi}
    {\rho_1 + \rho_2 + 2 \sqrt{\rho_1 \rho_2} \sin \varphi}
  + \frac{\hbar}{m}
    \frac{\left( \rho_1^{1/2} \nabla \rho_2^{1/2}
  - \rho_2^{1/2} \nabla \rho_1^{1/2} \right) \sin \varphi}
    {\rho_1 + \rho_2 + 2 \sqrt{\rho_1 \rho_2} \cos \varphi} .
 \label{e22}
\ee
The crossed term in (\ref{e20}) and (\ref{e21}) arises from the
interference of the probability densities and current densities
associated with $\psi_1$ and $\psi_2$ after assuming the coherent
superposition (\ref{e19}).
However, as seen from (\ref{e22}), the local velocity field keeps
a more complicated form, which cannot be expressed as a simple
superposition, thus indicating that the dynamics is going to be far
more complex.
This is illustrated in Fig.~\ref{fig1}, where the single wave-packet
Bohmian trajectories (red) diverge from the behavior displayed by the
two wave-packet ones as they approach interference region, ending up
eventually in the wrong place.
This can be readily understood by looking at the three graphs displayed
in Fig.~\ref{fig2}: although the probability density (part (a))
does not provide any clue, both the phase field (part (b)) and the
local velocity field (part (c)) immediately make apparent the dynamical
consequences of the superposition.

It is worth stressing that the non-crossing takes place regardless of
the presence of the quantum potential.
Rather it is an intrinsic or inherent property associated with the
dynamics generated through a guidance condition.
In classical mechanics we can also observe something similar, but in
phase space, where phase-space orbits cannot cross through the same
point at the same time.
In such a case, the ``guiding field'' is the classical action (this
was actually the idea that led Schr\"odinger in the derivation of his
wave theory of quanta \cite{sanz-bk-1}).
Nevertheless, in the two wave-packet problem presented above,
the non-crossing can still be well represented by the action of an
effective potential acting on a single wave packet, with the form of
a repulsive wall and an attractive region, both depending on time
\cite{sanz:JPA:2008}.

%%%%%%%%%%%%%%%%%%%%%%%%%%%%%%%%%%%%%%%%%%%%%%%%%%%%%%%%%%%%%%%%%%%%%%%

\subsection{Quantum probability tubes}
\label{sec2-3}

The possibility to study quantum processes in terms of Bohmian
trajectories brings in another interesting property, namely the
possibility to define {\it quantum probability tubes}
\cite{sanz:arxiv:2012}, i.e., tubes
in configuration space along which the integral of the probability
density at a time $t$ remains constant regardless of changes in their
profile or in the shape of the enclosed sections of $\rho$.
In order to show this, consider the so-called {\it restricted
probability}, defined as
\be
 \mathcal{P}(t) \equiv \int_\Omega \rho({\bf r},t) d{\bf r} ,
 \label{prob}
\ee
which gives the probability to find the system inside a certain region
of interest $\Omega$ of the corresponding configuration space at a
time $t$ \cite{sanz:cpl:2009,sanz:cpl:2010E,sanz:CP:2011}.
From standard quantum mechanics (and making use of the divergence or
Gauss-Ostrogradsky theorem), the variation with time of
$\mathcal{P}(t)$ can be expressed as
\be
 \frac{d\mathcal{P}(t)}{dt}
  = \int_\Omega \frac{\partial \rho}{\partial t}\ d{\bf r}
  = - \int_\Omega \left( \nabla \cdot {\bf J} \right) d{\bf r}
  = - \int_\Sigma {\bf J} \cdot d{\bf S} ,
 \label{ec5}
\ee
where $\Sigma$ is the boundary of $\Omega$.
That is, the losses/gains of $\mathcal{P}(t)$ inside $\Omega$ are
described by the outgoing/ingoing probability flow through $\Sigma$,
which is accounted for by the quantum current density ${\bf J}$.
Defining an arbitrary region $\Omega$ in standard quantum mechanics is
rather simple.
However, how can this region be propagated in time so that
$\mathcal{P}(t)$ keeps the same value along time?

Although the answer to the question posed above can be quickly answered
in classical mechanics by using arguments based on the Liouvillian
structure of this theory \cite{gutzwiller:1990}, the same cannot be
found in standard quantum mechanics, but in Bohmian mechanics and, more
specifically, in its non-crossing property.
As shown elsewhere \cite{sanz:arxiv:2012}, if the surface $\Sigma$ is
the geometric place formed by a (discrete or continuous) set of initial
conditions, because of the non-crossing property, any probability
enclosed inside such a set will remain constant at any subsequent time.
This is based on the fact that at any given time $t$ no trajectory
inside or outside the boundary defined by the set of {\it separatrix}
trajectories will be able to penetrate into or exit from $\Omega$
respectively.
Therefore, if the probability $P(t)$ is measured in terms of the
accumulation of trajectories embraced by $\Sigma$, i.e.,
\be
 \mathcal{P}(t) \approx \sum_{i=1}^N \delta({\bf r}_i(t) \in \Sigma(t)) ,
 \label{prob-2}
\ee
this number must remain constant at any time provided $\Sigma(t)$ is
a causal map of $\Sigma(0)$ at time $t$.
An important consequence of this result is that {\it any restricted
probability can be determined directly from the initial state if the
end points of the associated separatrix trajectories are known}.
This means that, in principle, one could determine (or, at least,
estimate) final probabilities without carrying out the full calculation
\cite{sanz:JPA:2011}, but directly from the particular region covered
by the initial wave function causally connected with the feature of
interest (e.g., a transmitted amplitude or a diffraction peak), since
\be
 \mathcal{P}_\infty
 = \int_{\Sigma_\infty} \rho ({\bf r}_\infty) d{\bf r}_\infty
 = \lim_{t \to \infty} \int_{{\bf r}(t)} \rho [{\bf r}(t)] d{\bf r}(t)
 = \lim_{t \to \infty} \mathcal{P}(t)
 = \int_{{\bf r}_0} \rho[{\bf r}_0] d{\bf r}_0 = \mathcal{P}_0 .
 \label{ec7}
\ee
Here, $\rho ({\bf r}_\infty) d{\bf r}_\infty$ is the probability to
find the system confined within the layer of configuration space
defined by the separatrix trajectories ending at ${\bf r}_\infty$ and
those ending at ${\bf r}'_\infty = [{\bf r} + d{\bf r}]_\infty$.
Analogously, $\rho [{\bf r}(t)] d{\bf r}(t)$ is the probability to find
the system confined within the layer defined by the sets of separatrix
trajectories ${\bf r}(t)$ and ${\bf r}'(t) = [{\bf r} + d{\bf r}](t)$
at a time $t$.
Note that, since the probability density is being evaluated along
quantum trajectories, it is preferably denoted as $\rho[{\bf r}(t)]$
instead of $\rho({\bf r},t)$, with ${\bf r}_\infty = {\bf r}(t \to
\infty)$.

\begin{figure}[t]
 \centering
 \begin{minipage}{21.5pc}
  \includegraphics[width=10pc]{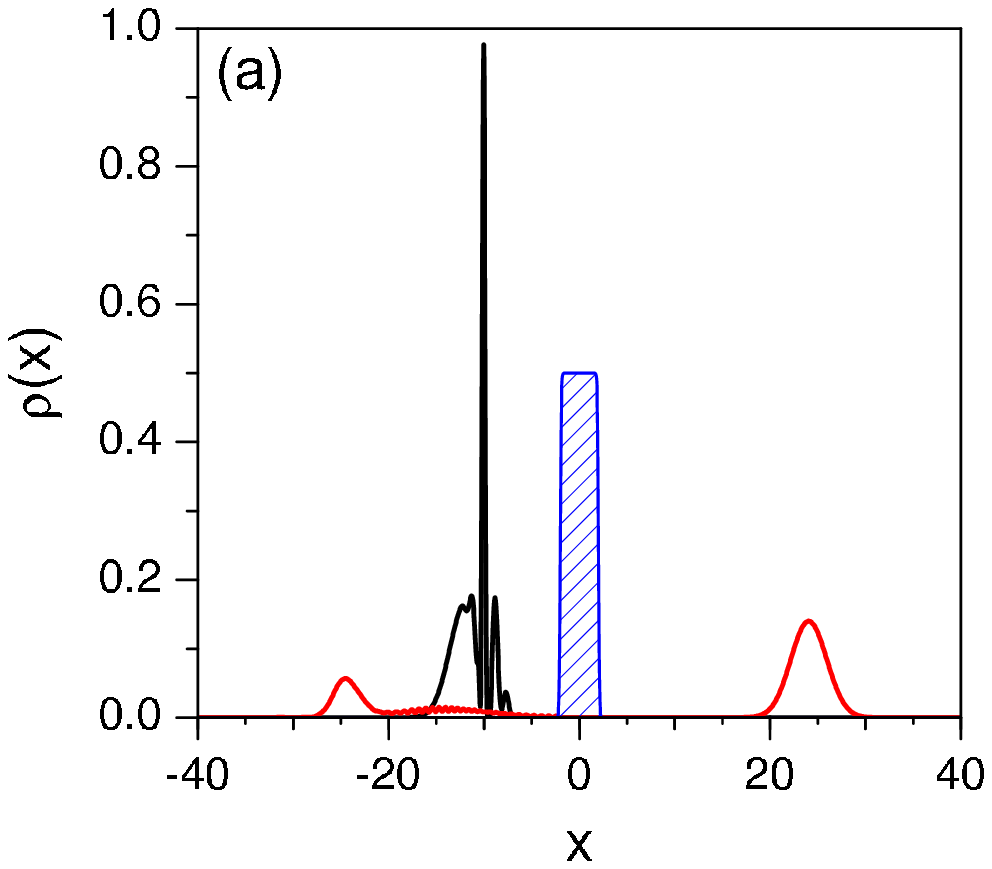}\hspace{1pc}
  \includegraphics[width=10pc]{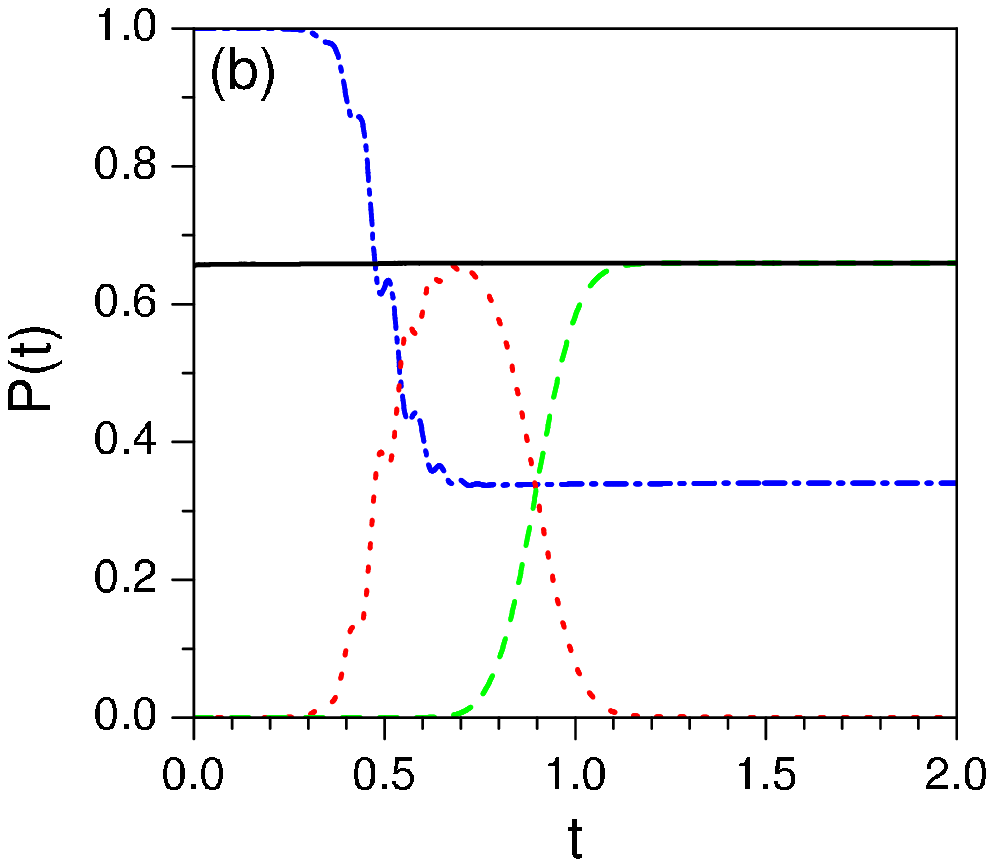}
  \includegraphics[width=10pc]{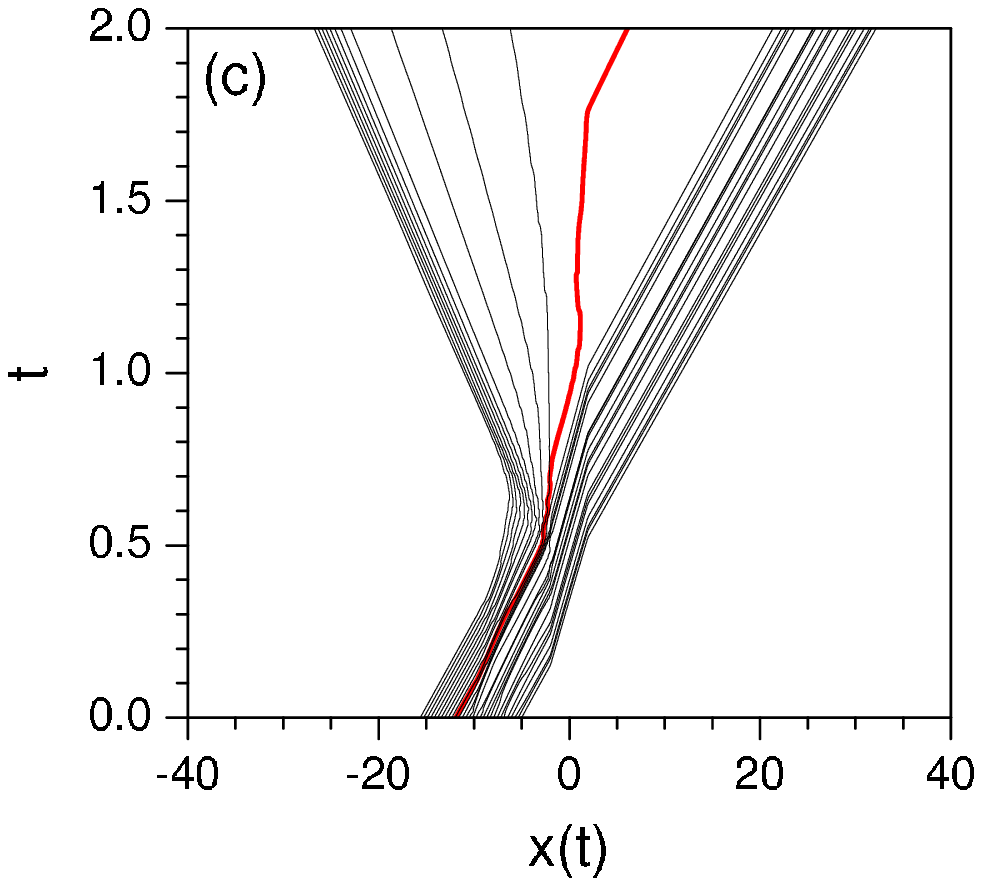}\hspace{1pc}
  \includegraphics[width=10pc]{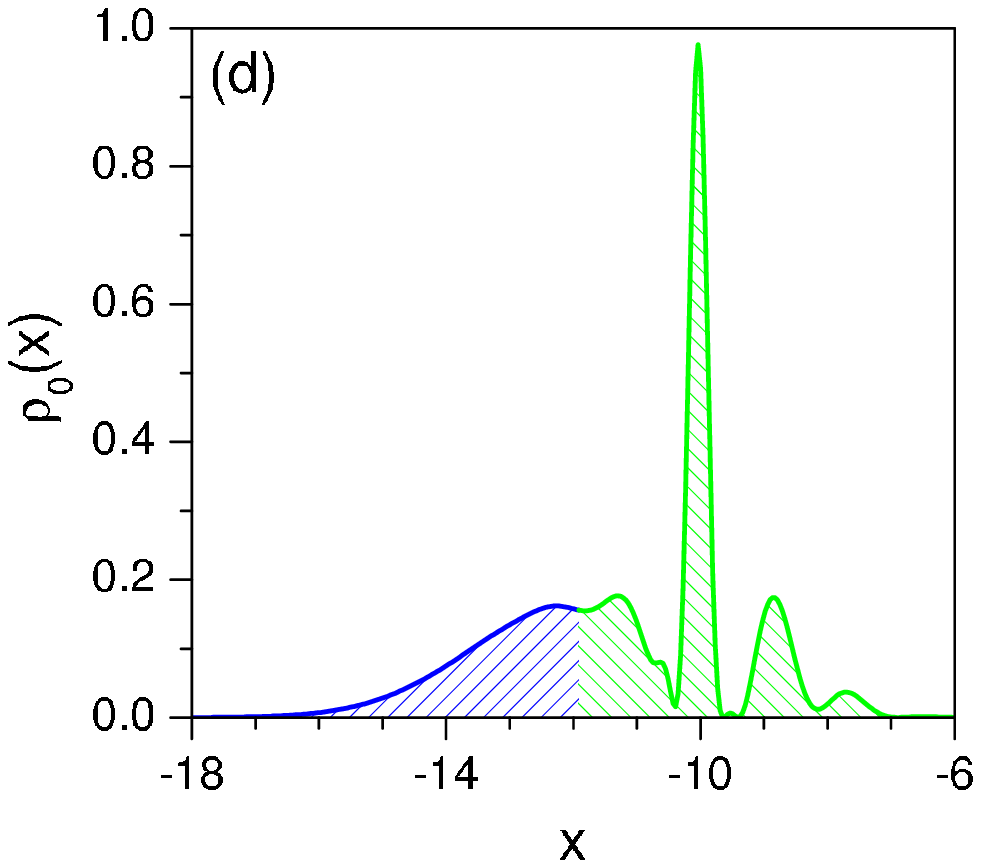}
 \end{minipage}
 \caption{\label{fig3}
  Scattering of a general wave packet by an almost square potential
  barrier.
  (a) Initial (black line) and final (red line) probability densities;
  the barrier is denoted with the blue shadowed region.
  (b) Time-dependence of different restricted probabilities:
  transmission (green dashed line), reflection (blue dash-dotted line)
  and intra-barrier resonance (red dotted line) probabilities.
  The black solid line denotes the transmission probability directly
  computed from the initial state (see text for details).
  (c) Bohmian trajectories illustrating the tunneling dynamics.
  The separatrix trajectory is displayed with the red thicker line.
  (d) Initial probability density split according to the separatrix
  initial condition: initial conditions in the green shadowed region
  contributes to transmission and those from the blue shadowed one to
  reflection.}
\end{figure}

In order to illustrate this result, consider the scattering of a wave
packet with a rather arbitrary shape by a barrier, as displayed in
Fig.~\ref{fig3}(a) (details on this calculation can be found in
\cite{sanz:arxiv:2012}).
The initial and final probability densities are denoted by the black
and red solid lines, respectively, while the barrier is indicated by
the blue shadowed region.
The propagation goes on until the probability in the region covered by
the barrier (intra-barrier resonance) becomes negligible, which can be
better seen in Fig.~\ref{fig3}(b), where the transmission (green dashed
line), reflection (blue dash-dotted line) and intra-barrier resonance
(red dotted line) probabilities are recorded along time.
These are the three restricted probabilities considered in this example,
with $\Omega_{\rm trans}$ being the region beyond the right edge of the
barrier, $\Omega_{\rm res}$ the region bound between the two edges of
the barrier, and $\Omega_{\rm refl}$ the region to the left edge of the
barrier.
After $t \approx 1.15$, $\mathcal{P}_{\rm res} \approx 0$ and
$\mathcal{P}_{\rm trans}$ reaches its maximum, asymptotic value;
$\mathcal{P}_{\rm refl}$, on the contrary, reaches its minimum earlier,
at $t \approx 0.75$.
This means than $\mathcal{P}_{\infty, {\rm trans}}$ can be computed by
simply choosing as the separatrix trajectory a Bohmian trajectory that
reaches the left tail of the transmitted wave packet in part (a) and
propagating it backwards in time, as seen in part (c).
Integrating the probability density between the position of this
trajectory at each time and infinity (for practical purposes), it is
then found that $\mathcal{P}_{\infty, {\rm trans}}$ remains constant
all the way, as the black straight line shows in part (b).
Regarding the initial set (probability density), it is found (see part
(d)) that any initial condition started to the right of the separatrix
initial condition will get into the transmitted region, while if it
starts to its left, the trajectory becomes reflected.

%%%%%%%%%%%%%%%%%%%%%%%%%%%%%%%%%%%%%%%%%%%%%%%%%%%%%%%%%%%%%%%%%%%%%%%
%%%%%%%%%%%%%%%%%%%%%%%%%%%%%%%%%%%%%%%%%%%%%%%%%%%%%%%%%%%%%%%%%%%%%%%

\section{Where can this knowledge be transferred to?}
\label{sec3}

It is very common to separate quantum physics from other wave theories
based on the difference in the treatment of probability amplitudes with
respect to other types of waves (which are considered as perturbations
propagating through a given medium or vacuum).
However, the way how these wave theories operate is in essence the
same, for both include the same type of concepts (e.g., coherence,
interference, diffraction, tunneling, discreteness, etc.) and
principles (e.g., uncertainty, superposition).
Taking this into account, it is possible to establish a sort of
feedback regarding the understanding of both wave phenomena, in
general, and quantum phenomena, in particular, through the hydrodynamic
(Bohmian) approach considered above.

To briefly show how this transfer of knowledge can be done,
consider the case of electromagnetic fields, for example.
Bohmian mechanics was formerly formulated to describe massive
particles, resulting very appealing to explain event-to-event
experiments like those carried by Merli \etal \cite{pozzi:AJP:1976},
Tonomura {\it et al.} \cite{tonomura:ajp:1989}, or Shimizu \etal
\cite{shimizu:pra:1992}.
Now, there are also event-to-event experiments with light, such as
those performed by Dimitrova and Weis \cite{weis:AJP:2008}, which
claim for a treatment on equal footing.
As shown by Prosser \cite{prosser:ijtp:1976-1}, this is actually
possible by directly considering Maxwell's equations (which put
electromagnetism at the same level of Schr\"odinger's wave mechanics
\cite{scully-zubairy-bk}).
In this case, given an electromagnetic field\footnote{Time-independent
electromagnetic fields will be considered here for simplicity, although
time-dependent ones should be used, in general.} defined by an
electric field ${\bf E}({\bf r})$ and a magnetic field
${\bf H}({\bf r})$, the two key elements necessary to define the
corresponding streamlines are the electromagnetic energy density
and the Poynting vector (electromagnetic current density), i.e.,
\begin{eqnarray}
 U({\bf r}) & = & \frac{1}{4}
  \left[ \epsilon_0 {\bf E}({\bf r}) \cdot {\bf E}^*({\bf r})
   + \mu_0 {\bf H}({\bf r}) \cdot {\bf H}^*({\bf r}) \right] .
 \label{e30} \\
 {\bf S}({\bf r}) & = & \frac{1}{2}\
  {\rm Re} \left[ {\bf E}({\bf r}) \times {\bf H}^*({\bf r}) \right] ,
 \label{e29}
\end{eqnarray}
respectively.
Since the electromagnetic energy density is transported through space
in the form of the Poynting vector, a local velocity field can be
defined \cite{bornwolf-bk} in analogy to (\ref{e10b}), which reads as
\begin{equation}
 {\bf S}({\bf r}) = U({\bf r}) {\bf v} ,
\end{equation}
from which electromagnetic energy flow lines or photon paths are
obtained by integrating the equation
\begin{equation}
 \frac{d{\bf r}}{ds} = \frac{1}{c}
  \frac{{\bf S}({\bf r})}{U({\bf r})} ,
\end{equation}
along the arc-length coordinate $s$ (which can be referred to a
proper time $\tau = s/c$, with $c$ being the speed of light).
By means of this procedure, the interference of diffracted polarized
beams was studied \cite{sanz:AnnPhysPhoton:2010,sanz:JRLR:2010},
showing a good agreement with experiments carried out later on
to infer the average paths displayed by photons
\cite{kocsis:Science:2011}.

The previous discussion focussed on Maxwell's equations.
However, it is not necessary to stay at such a level in order to define
hydrodynamic streamlines.
As mentioned above, this can be done at any level where the elementary
descriptive tool is a wave equation.
Thus, consider, for example, the transmission of a light pulse through
a waveguide within the small-angle or paraxial approximation
\cite{bornwolf-bk}.
Assuming that the optical axis is oriented along the $z$-axis and the
electromagnetic field passing through the waveguide is time-harmonic,
the field can be approximated by a plane wave along the $z$-direction
modulated by a certain complex-valued amplitude,
\be
 \Phi({\bf r}) = \phi({\bf r}) e^{ik_z z} ,
 \label{par1}
\ee
where $k_z = n_0 k$, $k=2\pi/\lambda$, $\lambda$ is the light
wavelength in vacuum and $n_0$ is the bulk refractive index.
Substituting this expression into Helmholtz's equation,
\begin{equation}
 \nabla^2 \Phi + n^2 k^2 \Phi = 0 ,
 \label{ch7-e8}
\end{equation}
with $n$ being the refractive index inside the waveguide, one readily
finds
\be
 2ik_z\ \frac{\partial \phi}{\partial z}
  + \frac{\partial^2 \phi}{\partial z^2}
  = - \nabla_\perp^2 \phi + (k_z^2 - n^2 k^2) \phi ,
 \label{part2}
\ee
where $\nabla_\perp^2$ is the transverse part of the Laplacian
(for example, in Cartesian coordinates, it reads as $\nabla_\perp^2 =
\partial^2/\partial x^2 + \partial^2/\partial y^2$).
If, apart from the paraxial approximation, it is also assumed that the
slowly varying envelope approximation holds (i.e., the amplitude $\phi$
varies slowly in space compared to $2\pi/k_z$), the highest-order
derivative in $z$ can be neglected (i.e., $\partial^2 \phi/\partial z^2
\approx 0$) and Eq.~(\ref{part2}) can be recast as
\be
 i\ \frac{\partial \phi}{\partial z}
  = - \frac{\nabla_\perp^2 \phi}{2k_z}
    + \frac{k}{2n_0} \left( n_0^2 - n^2 \right) \phi .
 \label{part3}
\ee
This equation is, indeed, isomorphic to Schr\"odinger's one, with $z$
playing the role of the evolution parameter (rather than the time $t$)
and $k_z = n_0 k$ the role of an ``optical mass''.
Equation (\ref{part3}) has been used, for example, to study the design
of waveguides with optimal conditions of light transmission
\cite{pepe:JLightTech:1998,pepe:AppOpt:1999,pepe:AppOpt:2003}.
However, regarding our discussion here, it also brings the possibility
to define a family of {\it optical streamlines} if the pulse
$\phi({\bf r})$ is expressed in polar from, as $\phi({\bf R},z)
= \rho^{1/2}({\bf R},z) e^{iS({\bf R},z)}$, where ${\bf R}$ denotes
the vector perpendicular to the propagation direction ($z$).
In this case, the local velocity field reads as
\be
 {\bf v} = \frac{d{\bf R}}{dz} = \frac{\nabla_\perp S}{k_z} ,
\ee
which gives rise to such streamlines when integration over $z$ is
carried out.

\begin{figure}[t]
 \centering
 \begin{minipage}{34pc}
  \includegraphics[width=10pc]{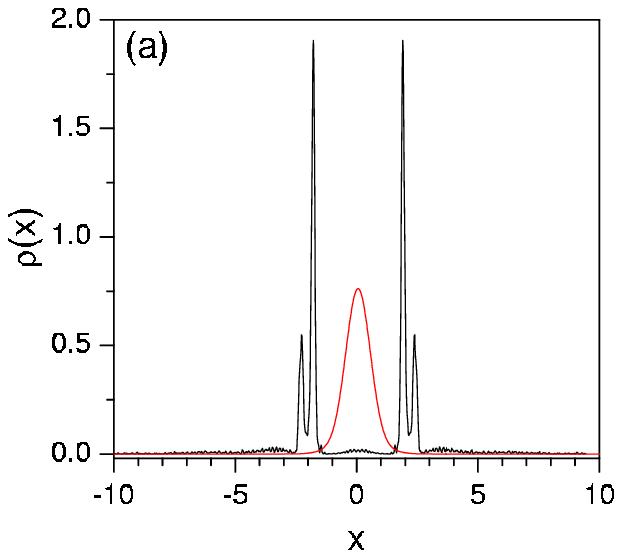}\hspace{1.5pc}
  \includegraphics[width=10pc]{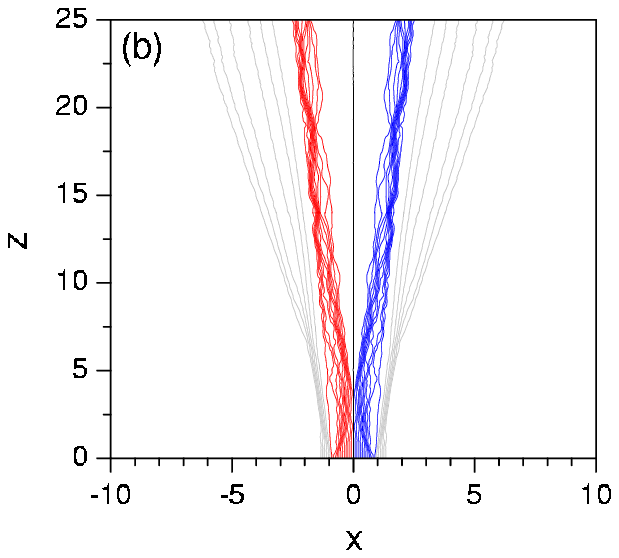}\hspace{1.5pc}
  \includegraphics[width=10pc]{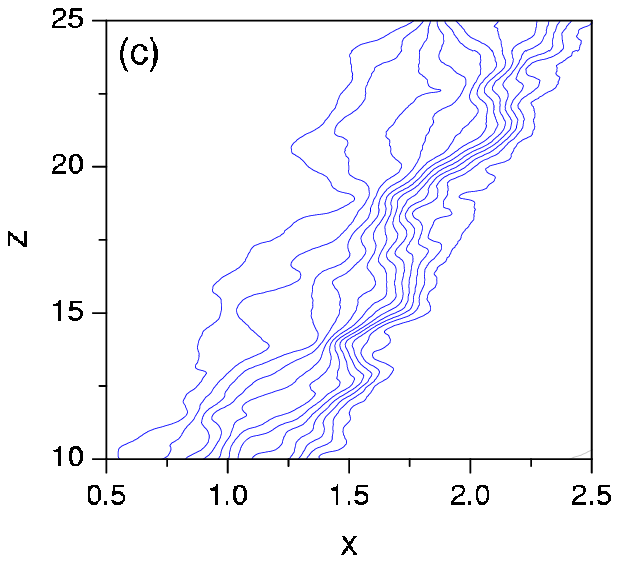}
 \end{minipage}
  \caption{\label{fig4}
   Light transport through a waveguide with the shape of a Y-junction.
   (a) Initial (black) and final (red) probability densities.
   (b) Optical streamlines illustrating the dynamics inside the
   waveguide.
   (c) Enlargement of part (b).}
\end{figure}

In Fig.~\ref{fig4} we have an example of how these trajectories look
like in the particular case of a Y-junction \cite{sanz:JOSA:2012},
which splits the initial pulse (red line) into two emerging pulses
(black line), as seen in part (a).
As it can be noticed in part (b), also here the electromagnetic flux
follows the non-crossing rule, which allows to distinguish between
different paths (and therefore to also define density tubes) without
disturbing the experiment, also in agreement with the experiment
performed by Kocsis \etal \cite{kocsis:Science:2011}.
Actually, in the enlargement of part (c), a behavior mimicking that of
the frustrated internal reflection (inside the waveguide) can be
observed, although contrary to geometric rays, the optical streamlines
follow the wavy behavior dictated by the pulse inside the wave, thus
being affected by diffractions and interferences.

%%%%%%%%%%%%%%%%%%%%%%%%%%%%%%%%%%%%%%%%%%%%%%%%%%%%%%%%%%%%%%%%%%%%%%%
%%%%%%%%%%%%%%%%%%%%%%%%%%%%%%%%%%%%%%%%%%%%%%%%%%%%%%%%%%%%%%%%%%%%%%%

\section{Final remarks}
\label{sec4}

We are used to consider that quantum mechanics is about small objects.
However, this is just a misunderstanding, for quantum mechanical
behaviors arise whenever objects are simple enough, i.e., whenever the
action of any internal or external surrounding degrees of freedom does
not play any relevant role in their dynamics.
As an example, just consider the interference patterns that can be
obtained from large organic molecules \cite{arndt:NatCommun:2011}
or Bose-Einstein condensates
\cite{ketterle:Science:1997,schumm:NaturePhys:2005}.
Such a misunderstanding comes from technological lacks in the past,
when quantum mechanics was developing, as well as {\it black-box}
interpretations, like the Copenhagen one.
The possibility to carry out interference diffraction of large objects,
particle by particle, requires of new viewpoints and interpretations
that may go beyond the blind ones available at present.

In this sense, the role of Bohmian mechanics as a theory of quantum
motion that describes the evolution of tracer particles when traveling
along quantum fluids (i.e., the fluid associated with the diffusion of
the wave function in a configuration space) are highly needed.
Even if at a statistical level predictions are the same as those
already provided by standard quantum mechanics, the fact that
hydrodynamic-like analyses can be carried out, elucidating properties
such as the flux non-crossing, results very appealing, for it allows
us to understand the (quantum) physical world in a different manner,
out of conceptual restrictions.

Furthermore, such approaches should also be extended to other fields
of physics, for they could help us to get different insights of already
well-known and established processes and phenomena.
This is the case, as it has been seen, of the so-called classical
electromagnetism.
In this case, not only theoretical descriptions have been developed to
better understand the evolution of the electromagnetic energy through
space, but also experiments have been motivated that have corroborated
the models proposed \cite{kocsis:Science:2011}.
Of course, and there is no doubt about the important role these
approaches present from a pedagogical viewpoint \cite{hermann:AJP:2002}
by allowing us to connect the distribution of energy (probability)
through space with its diffusion (current densities).

Finally, I would like to close this work with a simple, but striking
reflection.
Usually, one is always tempted to think that only those physical
problems that are directly connected with a particular practical
application are of interest and, therefore, worth spending time and
efforts to investigate them.
Of course, this is a legitimate point of view regarding where efforts
and resources should be targeted to.
However, this strategy cannot be used to get deeper into the physical
theories and approaches behind such problems in order to getter a
better and deeper understanding of them.
This is precisely what happens with quantum mechanics and, by extension,
with wave theories.
Nowadays nobody questions their use, but there are less and less
researchers facing the problem of the understanding of the quantum
world and the paradoxes it brings in to our intuition, which in spite
of all is {\it shaped} by a classical world.

%%%%%%%%%%%%%%%%%%%%%%%%%%%%%%%%%%%%%%%%%%%%%%%%%%%%%%%%%%%%%%%%%%%%%%%
%%%%%%%%%%%%%%%%%%%%%%%%%%%%%%%%%%%%%%%%%%%%%%%%%%%%%%%%%%%%%%%%%%%%%%%

\ack

The author would like to thank Gerhard Gr\"ossing for his kind
invitation to participate at the Conference ``Emergent Quantum
Mechanics'', within the 5th Heinz von Foerster Congress, held in
Vienna, as well as for his (and Sigfried Fussy, Johannes Mesa
Pascasio and Herbert Schwabl) hospitality and interesting discussions
during the event.
Moreover, he also wishes to acknowledge Profs. Basil Hiley and Helmut
Rauch for stimulating discussions and interesting questions.
This work has been supported by the Ministerio de Econom{\'\i}a y
Competitividad (Spain) under Projects FIS2010-18132, FIS2010-22082
and FIS2011-29596-C02-01, as well as by the COST Action MP1006
({\it Fundamental Problems in Quantum Physics}).
The author is also grateful to the Ministerio de Econom{\'\i}a y
Competitividad for a ``Ram\'on y Cajal'' Research Fellowship.

%%%%%%%%%%%%%%%%%%%%%%%%%%%%%%%%%%%%%%%%%%%%%%%%%%%%%%%%%%%%%%%%%%%%%%%
%%%%%%%%%%%%%%%%%%%%%%%%%%%%%%%%%%%%%%%%%%%%%%%%%%%%%%%%%%%%%%%%%%%%%%%

\section*{References}

%\bibliography{references}

\providecommand{\newblock}{}

\end{document}